\documentclass[final,5p,times,twocolumn,authoryear,a4paper]{elsarticle}
\usepackage{graphicx}
\usepackage{amssymb}
\usepackage{xcolor}
\usepackage{float} 

\journal{Digital Investigation}

\usepackage[hyphens]{url}
\usepackage[hidelinks]{hyperref}
\hypersetup{breaklinks=true}
\usepackage[font=scriptsize]{subcaption}
\begin{document}

\begin{frontmatter}


\title{\bf{A Forensic Audit of the Tor Browser Bundle}}


\author{Matt Muir, Petra Leimich and William J Buchanan}

\address{School of Computing, Edinburgh Napier University}

\begin{abstract}
The increasing use of encrypted data within file storage and in network communications leaves investigators with many challenges. One of the most challenging is the Tor protocol, as its main focus is to protect the privacy of the user, in both its local footprint within a host and over a network connection. The Tor browser, though, can leave behind digital artefacts which can be used by an investigator. This paper outlines an experimental methodology and provides results for evidence trails which can be used within real-life investigations.
\end{abstract}


\end{frontmatter}

\section{Introduction}

Use of the Tor Browser and network became mainstream in 2013, after the arrest and conviction of Ross Ulbricht - the alleged mastermind behind darknet marketplace “Silk Road” \citep{Dewey2013}. Prior to Ulbricht's arrest, a former American National Security Agency (NSA) contractor, Edward Snowden, had leaked classified files to The Guardian newspaper concerning the extent of mass surveillance by global intelligence services \citep{Greenwald2013}. This information, later known as “the Snowden revelations”, ignited a global debate surrounding the lack of privacy offered by digital communication. 

Motivated by fear of privacy-intrusion, members of the public and cyber criminals alike increasingly turned to the Tor Browser, tempted by its anonymising features. A new influx of users and increasing media attention catalysed academic research into the effectiveness of Tor and its ability to protect user privacy. The majority of this research concentrated on improvements to the performance and scalability of the network, as well as the investigation of remote attacks intended to unmask users \citep{Nordberg}. However, the question is whether anonymity is maintained where an adversary is local, rather than remote? 

This paper assesses the effectiveness of the Tor Browser in protecting the user against such an adversary by conducting a forensic analysis of the software and its interaction with the host operating system (OS). Our contribution is threefold:

\begin{itemize}
\item We show that artefacts proving installation/use of the browser are generated in memory and on disk in the form of default bookmarks. These artefacts are attributable to a particular user, uniquely identify the Tor Browser, and persist through uninstallation and logout. 
\item Furthermore, user activity within Tor is written to the Windows Registry as a consequence of recent updates to Windows 10. This allows a forensic adversary to determine the titles of pages visited using the browser.
\item From the results of our forensic audit, we devise a forensic methodology.
\end{itemize}

Our approach simulates typical web browsing activity with Tor. Using virtualisation and a pre-determined browsing protocol allows artefact recovery with static and live forensic techniques, such as process monitoring, keyword searching and file carving, with the aid of Autopsy and the Volatility Framework. 

Static analysis reveals significant leakage of user activity in the snapshots of machines used to perform the testing. This includes HTTP header information, web page titles and an instance of a URL. Further, live analysis identified traces of Tor processes even after the user had closed and uninstalled the browser and logged out. The absolute path to the browser executable was seen in RAM on several occasions, including the username of the user running the browser and the device from which it was run.

As a result, Tor is easily identified, cannot be securely deleted and activity from within a browsing session is determinable, suggesting that defence against a local adversary has not been fully addressed by the developers. This finding significantly expands on previous research, where artefacts pertaining to use of the browser, but not browsing history, were found.

\section{Background}
The novel approach to anonymous networking which Tor is (in)famous for has attracted interest from the cybersecurity community, largely resulting in research concerning de-anonymisation of users from a network perspective. However, it is the intention of the Tor Project to protect the user from both network and local adversaries. This is achieved through the implementation of features and design choices intended to obfuscate network activity and employ anti-forensics techniques to prevent browsing session data from being written to disk \citep{Perry2018}. 

\subsection{Tor Overview}

Tor, an acronym for \emph{The Onion Router}, refers to both a browser and a networking protocol developed by The Tor Project. The purpose of the Tor network is to mask the original source of Internet traffic by redirecting data from source to destination through a series of randomly chosen, volunteer-run servers \citep{Torproject.org}. Unlike regular web browsing, this obfuscates the user's IP address, ensuring that their identity is not disclosed to the websites they visit. By redirecting the traffic through encrypted channels, data is also protected from passive traffic analysis. 

Use of the network alone does not guarantee anonymity, making the Tor Browser Bundle (TBB) a necessary component in the privacy-oriented architecture \citep{Torproject.org}. The TBB is an extended support release (ESR) of Mozilla's Firefox browser, bundled with mandatory add-ons which protect user anonymity. The browser's design philosophy states that the TBB should protect the user from known web attacks designed to reveal a user's identity and also minimise the amount of browsing data written to disk \citep{Perry2018}. To enable this, plugins such as NoScript, which protects the user from malicious code on websites, are used by default, as is Firefox's Private Browsing mode. Forcing Private Browsing mode means that web history is not saved to disk by the browser itself, which should make it difficult for a forensic adversary to determine websites visited \citep{moz}. 

\subsection{Motivation}

Users of the Tor Browser and network rely on the perceived anonymity offered by the Tor Project. This research is motivated by the idea that little importance has been placed on protecting the user's anonymity from a forensic perspective, both by the Tor Project and by academic research in this area. If user anonymity is easily compromised by a forensic adversary, the browser could be considered a weak link in Tor's privacy-focused architecture. Whilst it is not currently illegal to use Tor in many states, shifts in attitudes towards the browser could take place as a reaction to political events, or as a wider drive for censorship. Moreover, users of Tor who see the browser as a tool for the greater good, such as journalists and political dissidents, may be easily identified if the browser leaves obvious traces of its use on computing systems. This could lead to persecution and further surveillance, regardless of legality. Understandably, the same features which make Tor attractive to the aforementioned users, also make it attractive to those who wish to cause harm. Therefore, a greater understanding of the forensic implications of using Tor is required by law enforcement investigating a range of digital crimes. It is this complex societal impact which has inspired the subject of this paper.   

\subsection{Traditional Browser Forensics}
Traditional web browsers record a significant amount of data from a browsing session - particularly in the default browsing mode. Web browser forensics involves the recovery of browsing artefacts which reveal information regarding a suspect's online activity. Browser forensics has become increasingly important for investigators as search history, download activity and page views can aid understanding of the criminal motive. 

Like other popular browsers, Firefox, the TBB's underlying browser engine, stores history, download and cookie information in SQLite databases, generally in clear text. This allows forensic investigators to easily retrieve browsing information \citep{Roussas2009, shepard, Noorulla2014}.
Acknowledging this lack of user privacy, browser developers introduced a private browsing mode, intended to limit or, ideally, eliminate browsing artefacts being written to disk. In a  comprehensive study, \citet{Montasari2015} found that the extent to which this aim has been achieved varies widely between browsers, with Firefox considered the second-most secure of the four mainstream browsers tested. 

Using a methodology which incorporated virtualised Windows platforms and the use of Autopsy to analyse snapshots, \citet{Montasari2015} attempted to recover cached web pages, web browsing history, download history, visited URLs and search terms from the target computer after performing a pre-defined browsing protocol using each browser's version of private browsing. A variety of typical activities, involving different types of web content, were covered:

\begin{enumerate}
\item Searching for and viewing a YouTube video.
\item Using Google to find a web page and then performing another search on the page.
\item Downloading a Facebook profile picture.
\item Searching and viewing an Amazon item. 
\item Searching for and viewing a PDF document using Google.
\end{enumerate}

The analysis was split into two stages; examining firstly “common” and “uncommon” locations on a hard drive and secondly the contents of the computer's volatile memory (RAM). Of the four browsers tested, Firefox was deemed relatively secure, although traces of search terms issued during the browsing protocol were discovered in pagefile.sys and the profile picture downloaded from Facebook was also discovered using file carving. 

These results confirm earlier work which found artefacts including search terms and URLs from private browsing sessions in pagefile.sys \citep{David2012, Findlay2014} and in RAM \citep{aggarwal2010analysis, Findlay2014}, and contradict Mozilla's claims that Firefox leaves no traces after termination of a private browsing session \citep{moz}.

\subsection{Overview of Tor Forensics Research}
Any weaknesses in Mozilla's implementation of private browsing suggest that the TBB could be susceptible to similar exploitation unless the developers have managed to mitigate these vulnerabilities. Acknowledging the need for a forensic audit, \citet{Sandvik2013} of the Tor Project conducted a cross-platform forensic analysis of the TBB on OSX, Linux and Windows. The TBB "aims to ensure that the user is able to completely and safely remove the bundle without leaving other traces on her computer" \citep{Sandvik2013}. However, the analysis results showed that the default behaviour of the operating system hosting the TBB could compromise user anonymity through its generation of artefacts related to features such as prefetching, paging and index searching, negating the anti-forensics measures implemented by the developers.

\citet{Darcie2014} found that analysis of the Windows Registry could determine the presence of the TBB after uninstallation, while \citet{Epifani2015} discovered during a real-life case study, where a rogue employee's laptop was seized, that Tor use on the target computer was proven by artefacts residing in Tor's install directory  as well as the behaviour of Windows prefetching. Hence the Tor Project either does not or cannot prevent a trail of evidence being created by the host OS. 

As the TBB was found to be successful in avoiding browsing data being written to the hard disk (besides in low-RAM situations) \citep{Darcie2014}, a static forensic analysis would likely yield evidence of installation and use. This could be advantageous to an investigator wishing to prove the use of Tor over other browsers, however, lacking artefacts of browsing history, it does not offer a picture of a subject's activity within Tor. A requirement for live forensic analysis became evident. 

\citet{Dayalamurthy2013} evaluated the work of previous forensic research using RAM dumps and suggested that this live forensics approach would be feasible when analysing the use of the TBB and attempting to de-anonymise Tor users. Her research proposed a forensic methodology which includes the use of the Volatility framework for memory analysis, along with the recovery of web graphics and analysis of the Windows Registry. However, this methodology was never tested.

\citet{Darcie2014} discovered that by creating a RAM dump of a target system, indexing it and searching for the keyword \emph{Tor}, numerous hits were returned in a computer used to download Tor, a computer where Tor was used and finally, a computer where Tor had been used and then uninstalled. This method of performing string searches on RAM captures echoes the methodology used by \citet{Montasari2015}. \citet{Warren2017} later expanded on this analysis of RAM by using the Volatility Framework, with the addition of community-developed plugins, to conduct an analysis of the RAM of a Windows 10 VMWare virtual machine. In doing so, evidence of Tor was found through the identification of Tor DLLs, environment variables, SIDs and command line usage.

\section{Methodology}
\begin{figure*}[h]
  \center\includegraphics[width=0.8\linewidth]{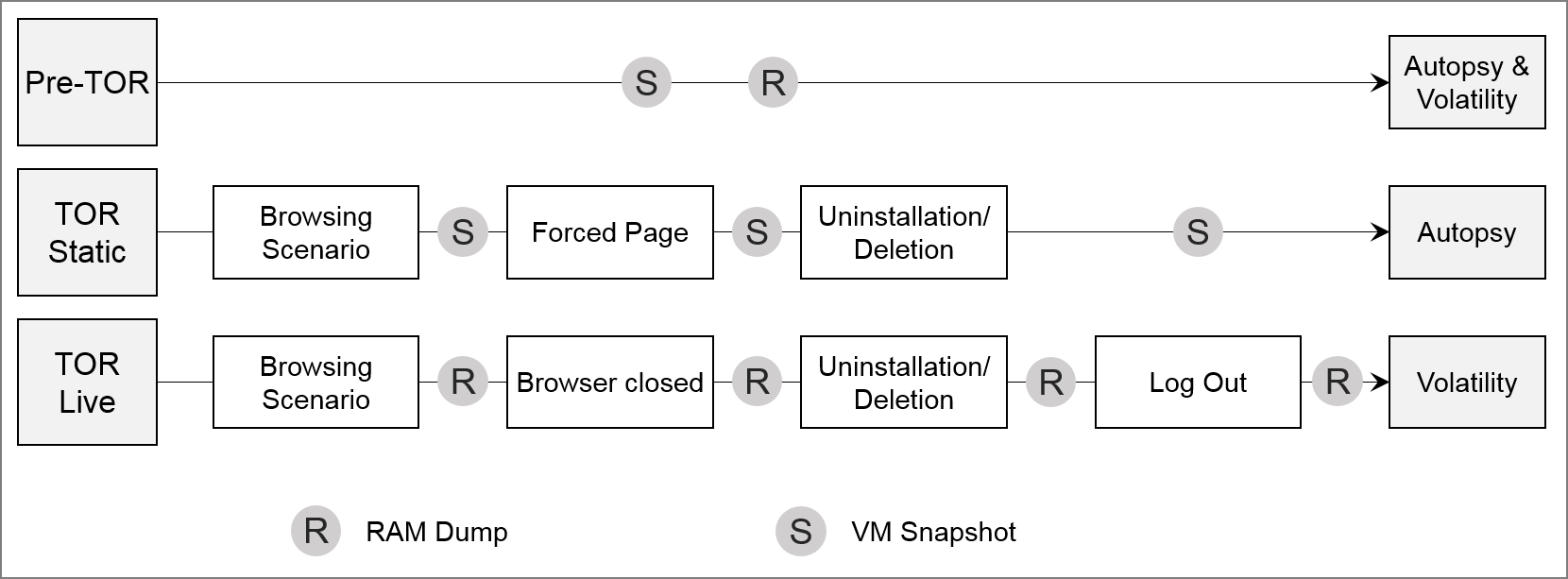}
  \caption{Overview of the Experimental Methodology}
  \label{fig01}
\end{figure*}

\subsection{Design considerations} \label{design}
Throughout this project, forensic analysis is intended to have two clear aims:

\begin{enumerate}
\item Expanding on previous research to confirm that the existence and use of the Tor Browser is still determinable in Windows 10, and
\item Proving that artefacts from the Tor browsing protocol can be recovered from the target computer.
\end{enumerate}

Live analysis of an application during runtime is particularly beneficial in order to understand how the host operating system and application interact. As the design efforts of the Tor project have focused on preventing writes to disk \citep{Perry2018}, a live analysis could potentially yield more information pertaining to the browsing session. Using an earlier version of Tor (3.6.1), \citet{Darcie2014} found evidence of web browsing in the form of JPEG and HTML files in live forensics but dead-box (static) forensics was unsuccessful. In a previous live forensics analysis of the Firefox browser, artefacts from a private browsing session were recovered from memory whilst the browsing session was open, and - to a lesser extent – after the browsing session had closed \citep{Findlay2014}. This showed that Firefox was able to terminate running processes, effectively flushing memory of artefacts of the browsing protocol when the user closed the Private Browsing window. However, whether or not this is also true for the TBB has not been established; this will be taken into account in our methodology.

To build upon previous research, our approach has been designed to answer the following questions:
\begin{itemize}
\item Does Tor manage to protect the user by flushing evidence of its use from RAM when the browsing session is closed?
\item Can Tor use be detected at four key moments; whilst the browser window is open, after closing the browsing window, after deleting the installation directory/associated files, and after the user has logged out?
\item Are files from the browsing protocol still recoverable in live forensics using Tor 7.5.2, the version current at the time of writing? 
\end{itemize}

Answers to the above should protect the user's anonymity from an adversary who manages to seize a personal computer during or shortly after Tor use, as well as after the seizure of a shared computer. Timing is a key parameter when designing a live forensics methodology. Figure \ref{fig01} gives an overview of the experimental methodology used in this project, clearly showing the timing of RAM captures for live forensics and virtual machine snapshots for static forensics. The experiments were repeated using Tor in mobile mode, i.e. run from a thumb drive, echoing \citet{Findlay2014} who found that mobile browsing reduced the artefacts left behind on the host, and using private browsing mode in native Firefox as a control.

Inspired by \citet{Montasari2015}, our test protocol mimics a user's browsing session with a variety of actions (i.e. searching, interacting with web forms, downloading, streaming media):
\begin{enumerate}
    \item Type theguardian.co.uk into the address bar and view the page.
    \item Go to https://support.mozilla.org and enter “profile folder” into the search box (submission to a web form).
    \item Go to images.google.com, search for “moog mother 32” and download the first image.
    \item Go to https://soundcloud.com/monolake/liveego1999 and stream the first three minutes of the embedded audio.
    \item At ebay.co.uk, search for “nike air force 1” and view the first item.
\end{enumerate}
The URLs and search terms were selected to include unusual strings to facilitate later keyword searches without false positives. 

\subsection{RAM Capture and Analysis Tools}
VirtualBox supports the exporting of the contents of RAM, provided the virtual machine is started from the command line with debugging mode enabled. The .pgmtophysfile filename command can then be issued at any point to dump the contents of RAM. The resulting ELF 64 file can then be analysed by the Volatility Framework (http://www.volatilityfoundation.org).  

Volatility was chosen due to its versatility as a command line tool, the wide range of plugins available for it and its effective use in previous Tor forensics research \citep{Warren2017}. With its vast array of plugins, Volatility is highly effective in analysing the state of memory at different points in time, building a comprehensive picture of the use of physical memory by an application. To address the aims stated in Section \ref{design}, the following Volatility plugins were chosen \citep{volFoundation}:

\begin{itemize}
\item cmdline: displays command line arguments for currently running processes.
\item dlllist: displays loaded DLLs for a given process.
\item dumpfiles: extracts memory-mapped and cached files.
\item envars: displays the environment variables used by processes.
\item pslist: prints all running processes by following the EPROCESS kernel structure.
\item pstree: prints the running processes as a tree, identifies parent and child processes.
\item shellbags: displays shellbag information taken from appropriate registry keys.
\item timeliner: creates a timeline from various artefacts in memory.
\end{itemize}

A common practice in live forensics is to observe running processes through the use of tools such as procmon \citep{Darcie2014}. Processes are assigned unique IDs (PIDs) in Windows and can be used to confirm that an application was running during the time of the analysis. Once the PID of a process is discovered, the process can then be searched in the output of Volatility's plugins, allowing an investigator to map its behaviour at certain points in time. Live analysis will take place first so that knowledge of the results which identify Tor use (for example, the names of DLLs) can then be used in a static analysis. Since it is considered likely that individually dumped RAM captures will expose contents of the browsing session (i.e. through caching), this was deemed an important inclusion to this project, particularly as \citet{Warren2017} was unable to do so due to time constraints.  Volatility's dumpfiles plugin will be used to achieve this and the resulting files analysed manually. This approach, inspired by \citet{Dayalamurthy2013}, should show changes to the Windows Registry, paths to the Tor install directory and contents of the browsing protocol.

\section{Results}\label{results}
\subsection{Live Forensics}

RAM was captured at four key moments in the experiment:

\begin{enumerate}
\item After the browsing scenario – with the browser window still open.
\item After closing the browser window.
\item After dragging the installation directory to the Recycle Bin and emptying (simulating an uninstall).
\item After the user logs out. 
\end{enumerate}

This should allow conclusions to be drawn which answer the questions asked in Section \ref{design}. As we found the artefacts left behind to decrease after each of the four stages, we present the results from the final stage.  

\begin{figure*}[h]
\begin{subfigure}{\textwidth}
  \centering
  \includegraphics[width=\linewidth]{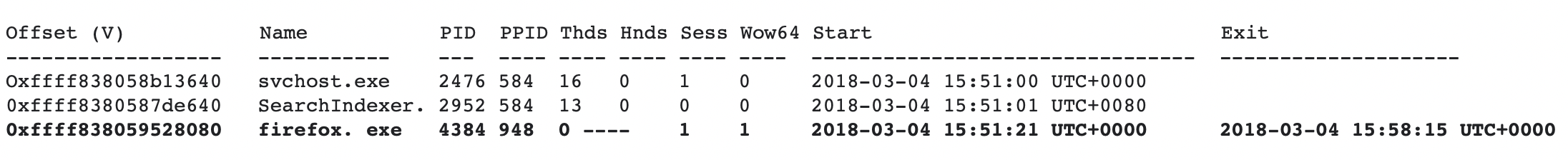}
  \caption{psscan}
  \label{fig02:sfig2}
\end{subfigure}
\begin{subfigure}{\textwidth}
  \centering
  \includegraphics[width=\linewidth]{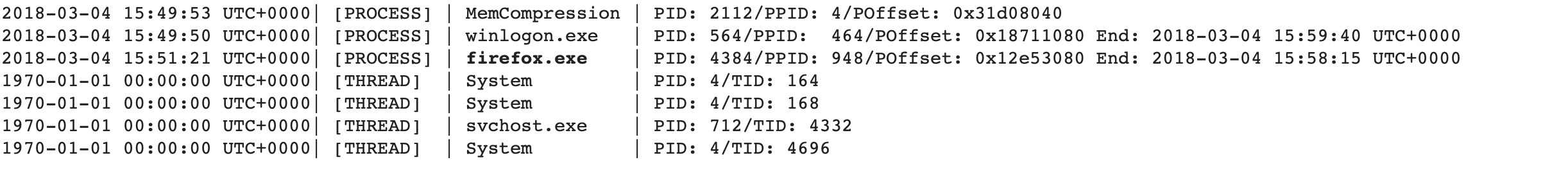}
  \caption{timeliner}
  \label{fig02:sfig4}
\end{subfigure}
\begin{subfigure}{\textwidth}
  \includegraphics[width=\linewidth]{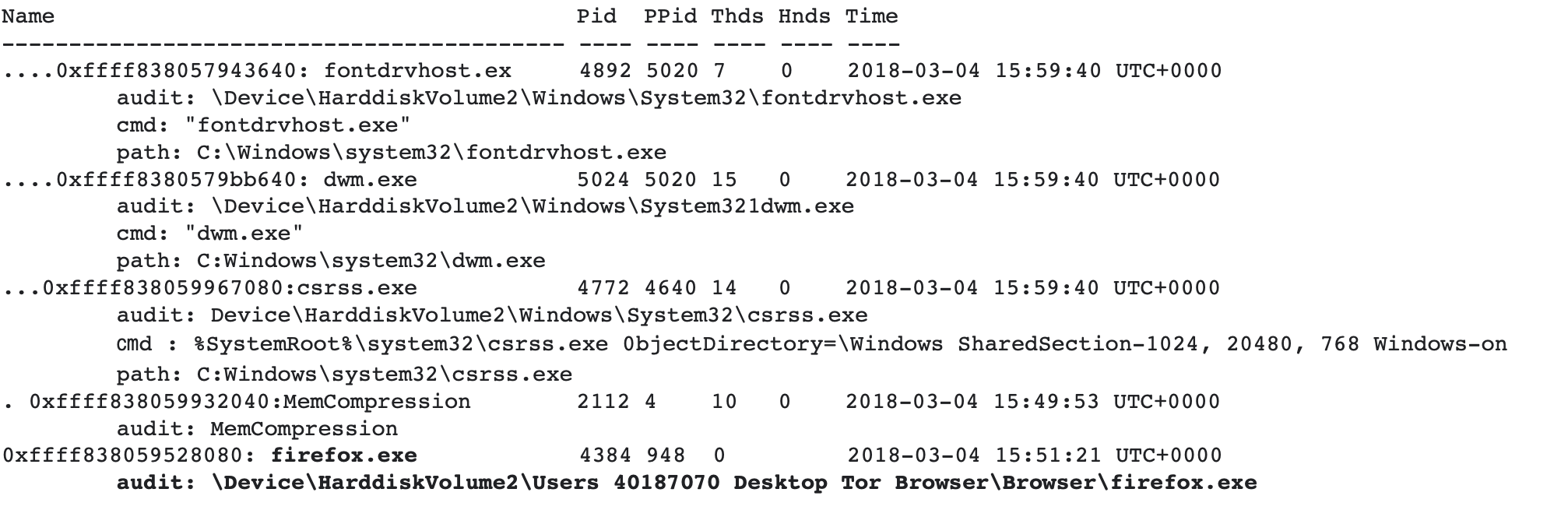}
  \caption{pstree}
  \label{fig02:sfig6}
\end{subfigure}  
\caption{Selected Volatility Plugin Outputs - After logout}
\label{fig02:fig}
\end{figure*}

After uninstalling the TBB and logging out, the Tor-related processes had been ended, but outputs from parsing the volatile memory with cmdline, pslist and dlllist plugins still showed the firefox.exe process with PID 4384; psscan and timeliner additionally showed timestamps (Figure \ref{fig02:fig}\subref{fig02:sfig2},\subref{fig02:sfig4}), while envars and shellbags produced nothing notable. While these findings could have been caused by a vanilla installation of Firefox, the output of the pstree plugin (Figure \ref{fig02:sfig6}) includes the absolute path to the Tor install directory, tying the process to the Tor browser. Via the PID and PPID, the start and end times from psscan and timeliner can then also be attributed to Tor.

\subsection{Static Forensics}
Snapshots of the VM were taken after three key stages in the experiment to enable a static forensic analysis to assess Tor's ability to prevent writing data to disk. Again, we focus on the final snapshot, taken after the uninstallation/deletion of Tor (Figure \ref{fig01}). 

\subsubsection{Keyword Searches}
\begin{figure}
  \includegraphics[width=\linewidth]{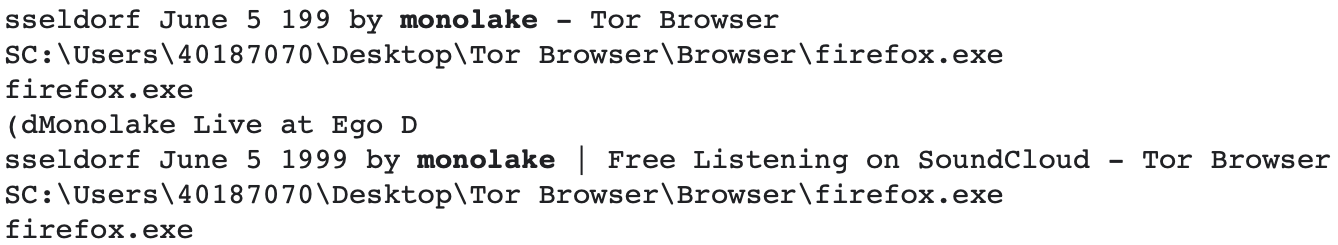}
  \caption{Results of \emph{monolake} string search, reflecting the title of the webpage (\textless title\textgreater Monolake Live at Ego D{\"u}sseldorf June 5 1999 by monolake \textbar  Free Listening on SoundCloud\textless/title\textgreater) suffixed with \emph{- Tor Browser}}
\label{fig07:fig}
\end{figure}   

The first step in the analysis was to carry out keyword searches for \emph{monolake}, \emph{soundcloud}, \emph{the guardian} and \emph{moog mother 32} to determine if and where these strings could be found in the forensic image. All search strings were found to appear in the same three files, with several occurrences in each file: NTUSER.DAT, ntuser.dat.LOG1 (a log of changes to NTUSER.DAT) and MEMORY.DMP. The only exception was \emph{moog mother 32}, which did not appear in ntuser.dat.LOG1 but instead appeared in unallocated space, suggesting that the log may have been flushed. Closer inspection of the contents of the three files revealed that all had exactly the same content, suggesting that the content of NTUSER.DAT was being logged and then replicated in MEMORY.DMP. Surprisingly, the file contents included the entire page title of each website visited suffixed with \emph{- Tor Browser} as well as the absolute path to the Tor install directory, clearly showing the username (40187070) as well as referencing firefox.exe within the Tor Browser directory. Figure \ref{fig07:fig} shows the \emph{monolake} results as an example.

\begin{figure}
  \includegraphics[width=\linewidth]{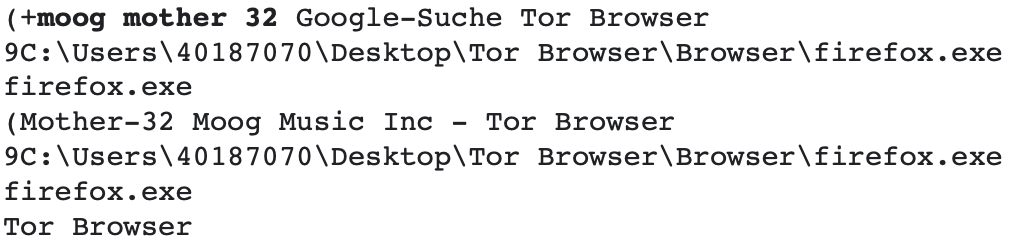}
  \caption{String search for \emph{moog mother 32} in NTUSER.DAT (extract)}
  \label{fig4_y}
\end{figure}  

The keyword matches for the string “moog mother 32” reflect its use in a Google search (Figure \ref{fig4_y}). Interestingly, the page title contained the German word for search, “Suche”, suggesting that the Tor exit node was located in a German-speaking country. The Downloads folder and explorer.exe also appear in this context, reflecting the saving of the downloaded image. As with the other keyword searches, the absolute path to the Tor directory is shown, with the Firefox executable clearly identified.

\subsubsection{Unallocated Space}
\begin{figure}
\begin{subfigure}{\linewidth} 
   \includegraphics[width=\linewidth]{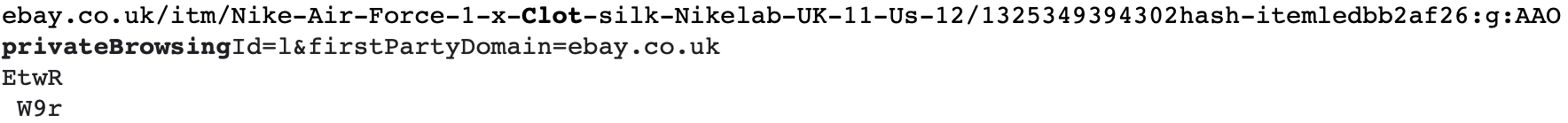}
   \caption{Ebay URL and Private Browsing evidence in .dll and .edb Files}
   \label{fig4:Sfigz}
\end{subfigure}
\begin{subfigure}{\linewidth} 
   \includegraphics[width=\linewidth]{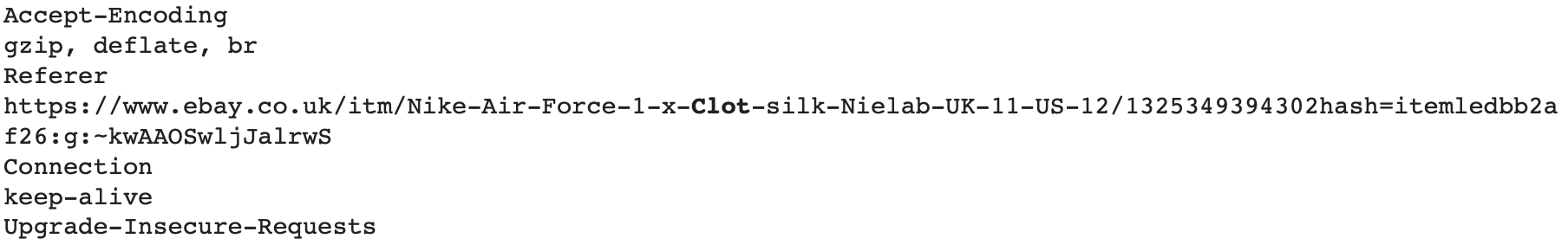}
   \caption{Ebay URL and HTTP Header Information in .reg Files}
   \label{fig4:Sfigw}
\end{subfigure}
\begin{subfigure}{\linewidth} 
   \includegraphics[width=\linewidth]{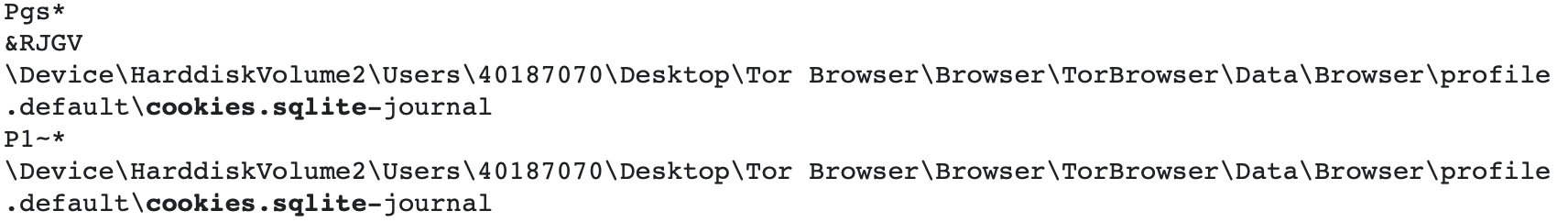}
   \caption{References to Tor Install Directory (extract)}
   \label{fig4:Sfig97}
\end{subfigure} 
\begin{subfigure}{\linewidth} 
   \includegraphics[width=\linewidth]{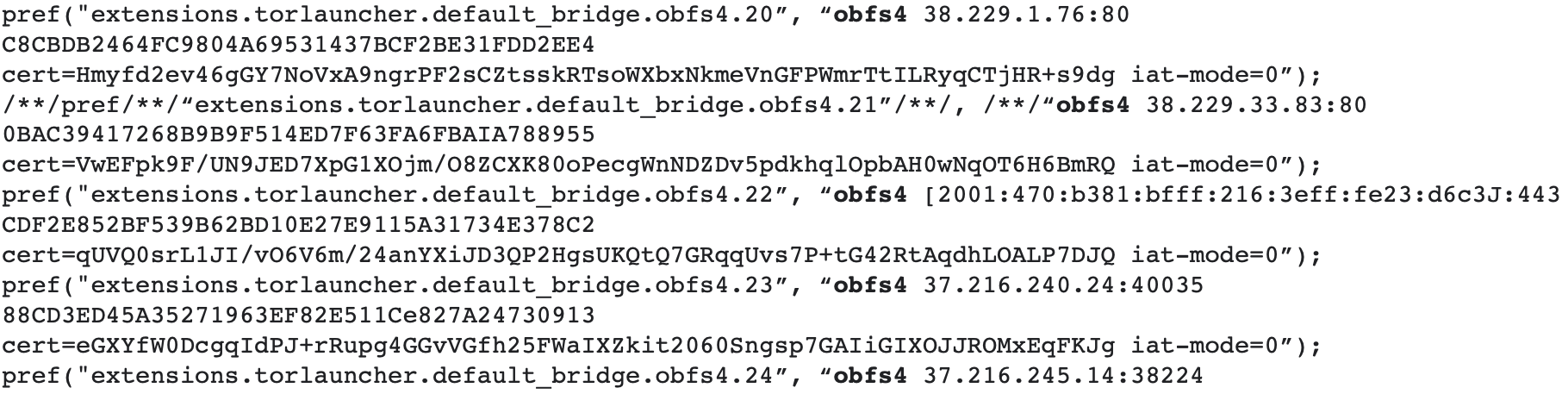}
   \caption{Obfs4 Bridging IP Addresses (extract)}
   \label{fig4:Sfig98}
\end{subfigure}  
\caption{Artefacts in Unallocated Space - After Delete}
\label{fig4}
\end{figure}

Several artefacts were recovered from unallocated space using Autopsy's carver module. When searching for the string 'clot' from the browsing protocol, six .dll, .edb and .reg files were discovered in unallocated space. Closer examination of these files showed considerable browsing data leakage: all contained the Ebay URL visited; the .dll and .edb files also referenced Private Browsing while the .reg file included remnants of HTTP header information (Figure \ref{fig4}\subref{fig4:Sfigz}-\subref{fig4:Sfigw}). Further searching of unallocated space uncovered references to the Tor installation directory and the obfs4 bridging IP addresses (Figure \ref{fig4}\subref{fig4:Sfig97}, \subref{fig4:Sfig98}). The browsing data found in NTUSER.DAT was also replicated in unallocated space. This may have occurred as a result of the machine crashing, however further analysis would need to be conducted to prove this. 

\subsection{Analysis of NTUSER.DAT}
As static analysis revealed an abundance of artefacts in the user profile's NTUSER.DAT registry hive, but Autopsy's parsing of registry is limited to string searches, NTUSER.DAT hives from several VM snapshots were extracted for further analysis.   

\begin{figure}
\begin{subfigure}{\linewidth} 
  \includegraphics[width=\linewidth]{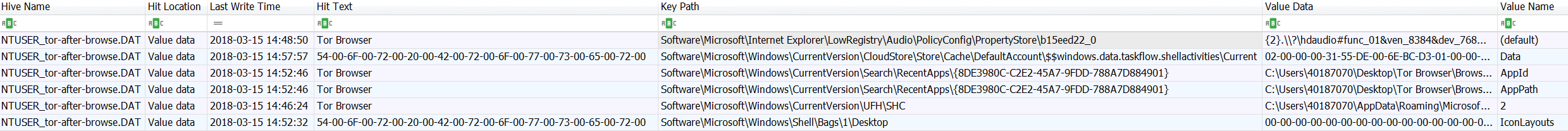}
  \caption{Tor, Standard Mode}
  \label{fig_keys:standard}
\end{subfigure}  
\begin{subfigure}{\linewidth} 
  \includegraphics[width=\linewidth]{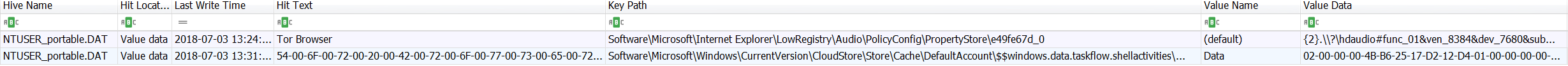}
  \caption{TOR, Portable Mode}
  \label{fig_keys:portable}
\end{subfigure}  
\begin{subfigure}{\linewidth} 
  \includegraphics[width=\linewidth]{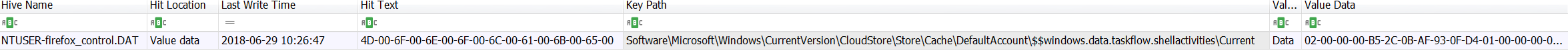}
  \caption{Firefox, Private Mode}
  \label{fig_keys:ff}
\end{subfigure} 
\caption{NTUSER.DAT Keys containing Search Strings}
\label{fig_keys}
\end{figure}

Firstly, a string search for 'Tor Browser' was carried out on the NTUSER.DAT file from the Tor After Browse snapshot with the free forensic tool Registry Explorer. Figure \ref{fig_keys:standard} shows the results of this search. Of particular interest is the Key Path column which shows the paths of the six different Registry keys where the string was found. 

\begin{figure}
 \begin{subfigure}{\linewidth} 
  \includegraphics[width=\linewidth]{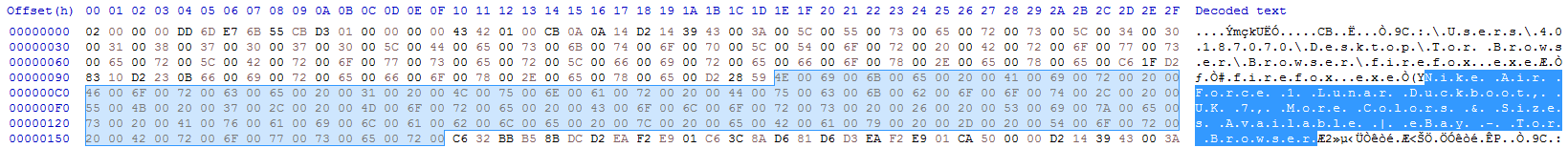}
  \caption{TOR, Standard Mode}
  \label{fig_shell_extracts:standard}
\end{subfigure}  
 \begin{subfigure}{\linewidth} 
  \includegraphics[width=\linewidth]{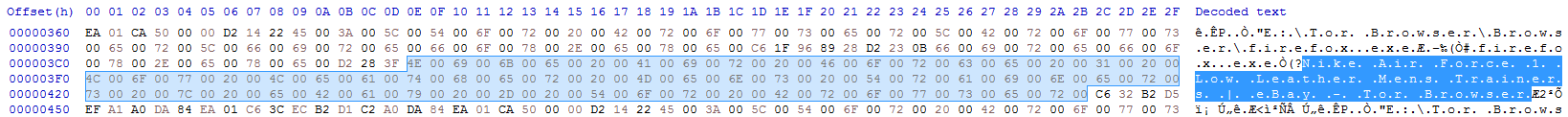}
  \caption{TOR, Portable Mode}
  \label{fig_shell_extracts:portable}
\end{subfigure} 
\begin{subfigure}{\linewidth} 
  \includegraphics[width=\linewidth]{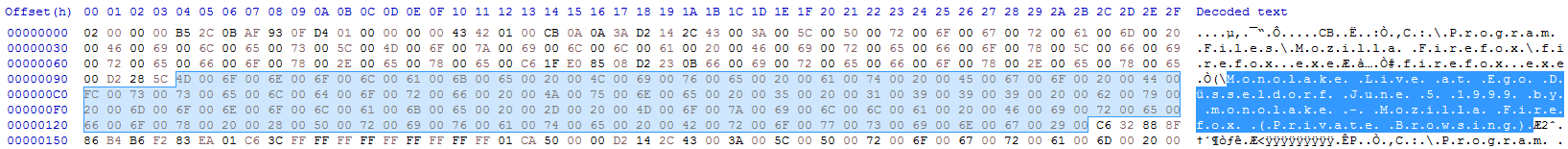}
  \caption{Firefox, Private Browsing}
  \label{fig_shell_extracts:ff}
\end{subfigure} 
\caption{Extracts of the shellactivities registry key showing titles of browsing tabs}
\label{fig_shell_extracts}
\end{figure}

The browsing data strings previously seen in Autopsy occur in HKEY\_CURRENT\_USER\textbackslash Software\textbackslash\nolinebreak[0]Microsoft\textbackslash\nolinebreak[0]Windows\textbackslash\nolinebreak[0]CurrentVersion\textbackslash\nolinebreak[0]CloudStore\textbackslash\nolinebreak[0]Store\textbackslash\nolinebreak[0]Cache\textbackslash\nolinebreak[0]DefaultAccount\textbackslash\nolinebreak[0]\$\$windows.\nolinebreak[0]data.\nolinebreak[0]taskflow.\nolinebreak[0]shellactivities\textbackslash Current (the \emph{shellactivities key} for short), making it the most interesting. The value, of REG\_BINARY type, contains page titles from the pages visited during the browsing protocol in Unicode as well as the absolute path to the application generating the browsing tabs, C:\textbackslash Users\textbackslash 40187070\textbackslash Desktop\textbackslash Tor Browser\textbackslash Browser\textbackslash firefox.exe (Figure \ref{fig_shell_extracts:standard}). 


To test whether writing of data to the Registry still occurs when the browser is run from an external device (in portable mode), the same analysis was performed on the NTUSER.DAT file extracted from a Tor Portable snapshot. While the string 'Tor Browser' was found in only two keys (Figure \ref{fig_keys:portable}), the shellactivities key again contained significant evidence from the browsing protocol including page titles and the absolute path to the Firefox executable within Tor's installation directory (Figure \ref{fig_shell_extracts:portable}). The latter, E:\textbackslash Tor Browser\textbackslash Browser\textbackslash firefox.exe, reflects the use of an external device, more details about which can be found by correlating the drive letter with registry keys such as USBSTOR.  

\begin{figure}
  \includegraphics[width=\linewidth]{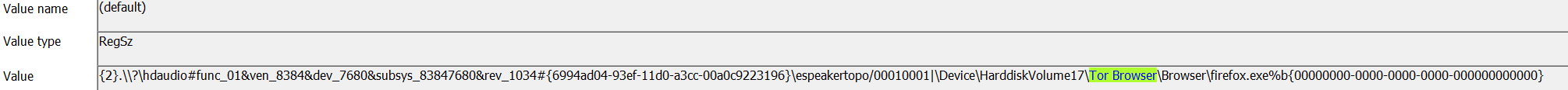}
  \caption{“Tor Browser” String in \textbackslash Audio\textbackslash PolicyConfig\textbackslash PropertyStore - TOR Portable}
  \label{fig4_102}
\end{figure}

In addition to the shellactivities key, the key HKCU\textbackslash Software\textbackslash\nolinebreak[0]Microsoft\textbackslash\nolinebreak[0]InternetExplorer\textbackslash\nolinebreak[0]LowRegistry\textbackslash\nolinebreak[0] Audio\textbackslash\nolinebreak[0]PolicyConfig\textbackslash\nolinebreak[0]PropertyStore\textbackslash\textless value\textgreater{} contained the 'Tor Browser' string in both the snapshot of Tor being run locally and in portable mode. Closer examination of the value data showed that it contains part of the path to the Firefox executable within the TOR browser installation directory (Figure \ref{fig4_102}). This key is likely used in the handling of audio by the host computer and the data would have been written to it when audio was streamed from the Internet as part of the browsing protocol. While the shellactivities key is a more comprehensive record of user activity, this audio key could corroborate the evidence, providing the user has used it to stream audio.

As a control, the same experiment was performed using the 'vanilla' Firefox browser in private browsing mode on a fresh VM, with a snapshot taken after executing the pre-determined browsing activities. As this control could not contain any references to the Tor browser, 'monolake' was used as the search string. The only match was in the shellactivities key (Figure \ref{fig_keys:ff}). Further inspection of this key in a hex editor revealed the same data leakage seen in the two Tor snapshots (Figure \ref{fig_shell_extracts:ff}).

\subsection{Decoding the shellactivities key contents}

It is evident that, under Windows 10, browsing data from user sessions is written to non-volatile storage, specifically the shellactivities registry key. This data leakage occurs regardless whether Tor, built upon the Extended Support Release (ESR) of Firefox, is used or Firefox's default Private Browsing mode, and also when a portable browser is used. This consistency suggests that the observed data leakage is a result of how the Operating System handles applications rather than some anomaly in the operation of Tor or Firefox. As a result, the anonymity-preserving intentions of both the Tor Project and Mozilla's private browsing mode are negated. Conversely, the shellactivities registry key is likely to contain valuable forensic artefacts and thus warrants further investigation. 

The parent of the shellactivities registry key, Cloudstore, does not exist in Windows 7, nor in the 1511 update of Windows 10. It first appears in the Windows 10 with the Anniversary Update, (1607 / 1603 in Win Education), but at that point it was not populated, having no subkeys or values. In Windows 10 machines running the Creator's Update 1703, the CloudStore key exists and has numerous subkeys that are populated with values, including the shellactivities key.

The data within the shellactivities is of REG\_BINARY type, i.e. hexadecimal. It appears to have an internal structure, which warrants decoding fully as this would allow plugins for forensic tools such as Regripper, Encase etc to be created. 

\begin{figure}[h]
  \includegraphics[width=\linewidth]{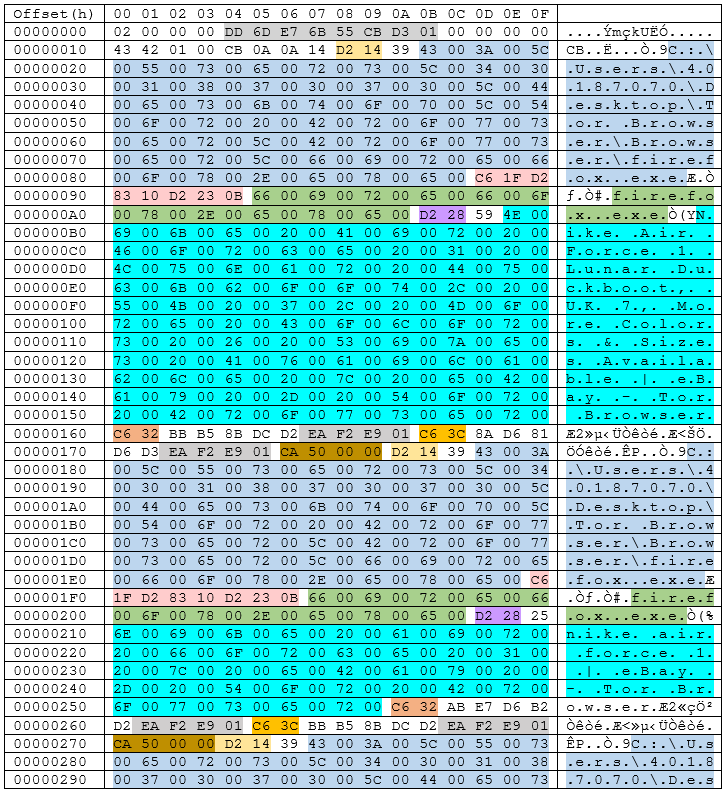}
  \caption{Internal structure of the shellactivities Key contents}
  \label{fig4_103}
\end{figure}

In Figure \ref{fig4_103}, we have colour-coded the start of the contents of this key after TOR was run in standard mode, uninstalled and the user logged out, in order to expose the internal structure.

\begin{itemize}   
\item The key header starts with 0x02000000; this is followed by an eight-byte Windows Filetime Timestamp (highlighted grey, here representing 3 April 2018 14:09:47), then 4 bytes of zero and 8 bytes of values (here 0x43420100CB0A0A14).
\item The first record entry starts at offset 0x18.
\item Every record starts with a record header signature of 0xD214 (buff), followed by one byte (here consistently 0x39, but varies in other examples; this could be a type, length or offset indicator of some sort).
\item This is followed by the absolute path to executable (blue-grey, variable length), then 8 bytes which are consistently 0xC61FD28310D2230B (pink), the name of the executable (green, variable length), 0xD228 (purple), 1 variable byte.
\item After the above, Page title (turquoise) + 0xC632 (orange) + 5 bytes + 0xEAF2E901 + 0xC63C (yellow) + 5 bytes + 0xEAF2E901 + 0xCA500000 (dark gold).
\item The underlined could be 1 byte + 8 byte timestamp of some sort - but see conversion attempts above.
\item This structure looks to be consistent for the actual web pages viewed, but the final few records vary somewhat.
\item If this is correct, the individual entries are delimited by 0xD21439 header and 0xCA500000 footer, and internal components separated by specific values as well, so size of entry would not need to be stored.
\item This entry header and footer might then be searchable, e.g in registry entries, or possibly unallocated space if deleted? (but I don't know how registry deals with deletion - possibly internally).
\end{itemize}

\section{Findings}

\subsection{Proposed Forensic Methodology}
The interactions between Tor and Windows described in Section \ref{results} can be used to design a generalised forensic methodology for identifying the use of TOR and Web pages visited using TOR or Private Browsing.
\begin{itemize}
\item Where possible, take a RAM dump. Analyse with Volatility's psscan, pstree and timeliner plugins to (a) establish the use of TOR and (b) find the username. This will also reveal timestamps and can be carried out even after the user has uninstalled TOR and logged out. Where they exist, pagefile.sys or hiberfile.sys can be used instead of a RAM dump.  The analysis of all three of these data sources from the same system could result in the recovery of different, but nonetheless relevant - and corroborative or complimentary - evidence.
\item From non-volatile storage, extract the NTUSER.DAT registry hive of the previously identified user, or, failing this, of all users.
\item Extract the contents of the shellactivities key. Search for Tor and/or firefox to find the titles of web pages visited. Similarly, a search for other browser executables will reveal visited pages even if private browsing was used.
\item A keyword search for 'obfs4' in unallocated space can reveal bridging IP addresses that may have been used by TOR.
\end{itemize}

\subsection{Live Forensics}
The live forensics methodology in this project showed that by using the Volatility Framework and associated plugins, four processes which can be attributed to Tor are created at runtime. These processes are as follows:

\begin{itemize}
\item firefox.exe: two occurrences of this process were seen.
\item tor.exe.
\item obfs4proxy.exe: it is unlikely this would have appeared had the obfs4 proxy not been used.
\end{itemize}

Whilst the names of these processes are obvious, the output of several of the Volatility Framework plugins also revealed the absolute path to the installation directory where the browser executable resides - providing further indication of Tor use. In doing so, the disk from which the browser was run, the username of the user running the browser, and the name of the browser can be clearly seen. This means that by using the Volatility Framework on a RAM dump seized whilst the browsing session is still active, a forensic adversary can positively identify the use of Tor.

After termination of the session, The Tor Browser manages to end tor.exe, obfs4proxy.exe and one of the firefox.exe processes. This makes it significantly more difficult for a forensic adversary to confirm that Tor was running on the machine at the time of seizure and coincides with the behaviour of Firefox observed by \citet{Findlay2014}. Despite this, one firefox.exe process persisted. Although the process can still be seen, the environment variables, command line arguments, and DLLs which were attributable to this and the other three processes could no longer be identified after session termination. This suggests that the persisting firefox.exe process could not be fully terminated by closing the browser window and now exists in a traceable but inactive state. Regardless, the output of Volatility's pstree plugin still managed to attribute it to Tor by identifying the absolute path to the install directory. Without this finding, it would be impossible to determine that this process had originated from Tor and not a regular installation of Firefox. 

This behaviour continued even after the user logged-out, proving the pstree plugin to be invaluable when attempting to prove the use of Tor on a shared computer. 

In summary, Tor use can be easily detected using live forensics, particularly when the browsing session is still active. Ensuring that the browsing session is closed after use helps to conceal the fact that Tor was used. However, an artefact (firefox.exe) remains detectable in RAM after closure, deletion, and log-out. It is likely that the traceable artefact is the result of an anomaly in Firefox's handling of running processes. This belief is strengthened by the fact that Tor manages to remove all evidence of the processes directly attributed to its browser, yet one Firefox process remains. Perhaps this abnormality was introduced in an update of Firefox's Extended Support Release, or it may even be an unforeseen result of the interaction between Tor's plugins and the underlying browser. Nonetheless, it shows that reliance on a third-party browser can introduce problems which undermine user anonymity. 

To expand on this, an examination of the files dumped from the contents of RAM will need to be performed. This will determine whether evidence of the browsing protocol is also present in RAM, a significantly greater finding than the identification of running processes. Unfortunately, time-constraints made this impossible in this project, an evaluation of which follows in a later section. 

\subsection{Static Forensics}
\subsubsection{Browsing Data Leakage}
Perhaps the most interesting and unexpected conclusion of this project is that Tor does, in fact, write browsing data to disk. Artefacts from the browsing protocol, including HTTP header information, titles of web pages, and an instance of a typed URL were found in several Registry files. This means that use of static forensics by forensic investigators is potentially more worthwhile than examining the contents of RAM, at least for this version of the browser. \citet{Epifani2015} and \citet{Darcie2014} both mention the Registry as an area where Tor evidence could be recovered. In the Registry, \citet{Darcie2014} identified evidence of Tor presence, but not of browsing data. However, their searches were limited in only including the keyword “Tor”. Similarly, \citet{Epifani2015} showed that Tor could be identified in the NTUSER.DAT Registry hive, within the User Assist Key. Yet, they determined that only the number and time of execution could be found. Therefore, the findings of this research which show that the majority of the browsing protocol can be recovered from NTUSER.DAT are particularly significant. This suggests that in the versions between 4.0.2 \citep{Darcie2014} and 7.5.0, a design change occurred which has led to this data leakage.

The vast majority of the browsing protocol was found in the NTUSER.DAT Windows Registry file. This makes it possible for a forensic adversary to reconstruct the activities of the user in relation to the browser. It was unclear what caused this information to be written to that area of the Registry, however multiple.reg files which are used to add changes to the Registry were found in the snapshot taken after the browser had been deleted. These files contained an Ebay URL visited during the browsing protocol, making it likely that some aspect of Tor was writing the contents of memory to the Registry. This is backed-up by the discovery of the ntuser.dat.LOG1 file, this file logs changes to the NTUSER.DAT Registry file and contained the same browsing data-leakage.

\subsubsection{Firefox Artefacts}
The two SQLite databases used by Firefox to track cookies and history (cookies.sqlite \& places.sqlite) were both recoverable from the file system after deletion. Cookies.sqlite contained nothing of relevance but its existence could indicate that Tor was installed on the computer - if it could be proven that it did not come from a regular Firefox installation. Places.sqlite contained no information about the browsing protocol, however, it did contain entries for the Tor Project's blog and a Learn About Tor page. This seems counter-productive when designing a browser which is designed to be anonymous. The developers of the Tor Project may see this as a trade-off between security and usability, yet the fact that it can still be recovered after deletion is further proof that the browser is incapable of being securely deleted. Since the use of Tor is not currently illegal in most countries, mere identification of the browser may not be considered significant. However, these artefacts may have forensic value when trying to prove that Tor was used over another browser or as part of a wider investigation.   
\section{Key Findings}
This section summarizes some of the key findings.
\subsection{Live Forensics}
\begin{itemize}
\item Four processes can be attributed to Tor whilst the browser is open.
\item Use of the bundled obfs4 proxy (used to bridge connections to the Tor network) can be detected.
\item After the user terminates the browsing session, only one Firefox process remains.
\item Using the Volatility Framework, the path to the browser and the run time can be identified. Even after the user has closed the browsing window, deleted the browser and logged-out.
\item As a result of finding the absolute path to the browser, the username, location of the browser and use of Tor versus Firefox can be confirmed.
\end{itemize}

\subsection{Static Forensics}
The following are observations for static forensics:

\begin{itemize}
\item The design aim of preventing Tor from writing to disk \citep{Perry2018} is not achieved in this version.
\begin{itemize}
\item Configuration files, downloaded files, and browser-related data are recoverable from the file system.
\item Significant data-leakage from the browsing session occurred: HTTP header information, titles of web pages and an instance of a URL were found in registry files, system files, and unallocated space.
\item The data-leakage contained the German word for 'search' in reference to a Google search. This hints at the locale of the Tor server used to exit the network (exit relay).
\end{itemize}

\item The Tor Project's design aim of enabling secure deletion of the browser \citep{Sandvik2013} is not achieved in this version.
\begin{itemize}
\item References to: the installation directory, Firefox SQLite files, bridging IPs/ports, default bookmarks, Tor-related DLLs and Tor product information were all recovered after the browser was deleted.
\item In a scenario where the operating system paged memory, an instance of a URL from the browsing protocol along with references to private browsing was found. 
\item The file “state” used by Tor to track successful launches of the application was found after browser deletion. Prior to being deleted, this file contained the last run time, however, in the recovered file this timestamp was overwritten. This suggests that Tor manages to overwrite this data.
\end{itemize}
\end{itemize}

\subsection{Forensic Implications of Using Tor}
Acknowledging the above results, a forensic investigator could prove:
\begin{itemize}
\item The use of Tor even after a user has attempted to delete the browser.
\item The last execution time/date.
\item Evidence of pages visited during a browsing session.
\item The language of the country where the server used to exit the Tor network resides.
\item That Tor was used over Firefox.
\end{itemize}

\subsection{Limitations}
Some limitations in the methodology of this paper can be observed after completion. One such limitation is that the networking aspect of Tor was not used to aid the forensic examination. For example, \citet{Warren2017} uses Volatility's netscan plugin to observe open network sockets, determining the TCP ports used by Tor in live forensics. This could be used to further prove the use of Tor. Equally, time constraints in this project meant that the individual files dumped from memory could not be analysed. This is likely to have resulted in further evidence of the browsing protocol. Furthermore, to find the browsing data-leakage seen in this project, a knowledge of unique strings related to the browsing protocol was required. This would be unlikely in a real-life static analysis, although if it was known that data from Tor is leaked to areas of the Registry such as NTUSER.DAT, the same data-leakage could be found. Finally, the experiments performed suffered from a lack of control. With more time, the same browsing protocol and analyses could be performed on a system running Firefox. This would allow conclusions to be drawn about which artefacts were attributable to Tor and which were attributable to Firefox. From this, an assessment of whether Firefox's implementation of private browsing or Tor's privacy-enhancing mechanisms led to the data-leakage observed in the results could be made.

\section{Conclusions}
The limited research in the area of Tor Forensics discovered during the Literature Review suggested that a need for live forensics had become increasingly important. This was largely due to recent papers concluding that Tor did not write browsing data to disk \citep{Warren2017}. Prior to this, \citet{Darcie2014} had successfully managed to retrieve files residing in RAM from their browsing protocol, even after the browser was deleted. This coincided with the forensic methodology proposed by \citet{Dayalamurthy2013}, which also placed emphasis on live forensics.  A new methodology was proposed, which favoured a live forensic analysis followed by a static analysis of the virtualised test machine. This resulted in a more thorough forensic process, allowing the maximum level of knowledge about the browser to be gained in the limited timeframe available.

Although previous researchers had successfully recovered artefacts of Tor-use, a comprehensive analysis of both volatile and non-volatile memory had yet to be completed – with most of the existing research concentrating on a singular aspect of computer memory. Likewise, the idea of the Tor Browser being used in a shared-computing scenario had yet to be fully explored. \citet{Sandvik2013} had contemplated this, yet it seemed inadequately addressed in her results. Therefore, it became obvious that the design of an experiment which incorporated the action of a user logging-out should be devised. It was determined that in such a scenario, a forensic adversary would likely conduct live forensics on the shared computer \citep{Tsalis2017} and the purpose of this experiment would be to determine whether Tor could be identified running on the system and whether evidence from the browsing protocol could be extracted from live memory. 

Due to the volatile nature of RAM, acquisition of live memory is rarely possible in the field. This is applicable even in shared-computing environments, as often the user can power cycle a shared computer without consequence. Considering that the intended audience for this project was both users of the browser and forensic investigators wishing to analyse it, the omission of a subsequent static analysis would constitute an incomplete methodology. This is especially true as a large number of Tor users will likely use their personal computer which could be subject to seizure by a forensic adversary. Therefore, the multi-faceted experimental design was required. This proved successful in the end as many unexpected results were born from the static analysis, an aspect which may have been omitted if too much reliance had been placed on the results of previous research. 

The technique of indexing the hard drive and applying keyword searches based on known Tor artefacts and the browsing protocol was simple yet is something that the browser should protect against. In particular, \citet{Perry2018} state explicitly in their design philosophy that browsing data leakage should not happen. This indicates that the Tor Browser does not adequately protect the user from a forensic adversary. 

\bibliographystyle{model2-names}\biboptions{authoryear}
\bibliography{sample.bib}

\end{document}